\documentclass[conference]{IEEEtran}
\IEEEoverridecommandlockouts
\usepackage{cite}
\usepackage{amsmath,amssymb,amsfonts}
\usepackage{algorithm}
\usepackage{algorithmic}
\usepackage{graphicx}
\usepackage{textcomp}
\usepackage{xcolor}
\usepackage{balance}
\usepackage{siunitx}
\usepackage{caption}
\usepackage{subcaption}
\usepackage{multirow}
\usepackage{makecell}
\usepackage{optidef}
\usepackage[a4paper,
            left=0.75in,
            right=0.75in,
            top=0.7in,
            bottom=1in]{geometry}

\def\BibTeX{{\rm B\kern-.05em{\sc i\kern-.025em b}\kern-.08em
    T\kern-.1667em\lower.7ex\hbox{E}\kern-.125emX}}
\begin{document}

\title{Conditional Denoising Diffusion for ISAC Enhanced Channel Estimation in Cell-Free 6G
\vspace{-1ex}}

\author{
\normalsize
Mohammad Farzanullah\IEEEauthorrefmark{1}, Han Zhang\IEEEauthorrefmark{1}, Akram Bin Sediq\IEEEauthorrefmark{2}, Ali Afana\IEEEauthorrefmark{2} and Melike Erol-Kantarci\IEEEauthorrefmark{1}\\
\vspace{0ex}
\IEEEauthorblockA{\IEEEauthorrefmark{1}
    School of Electrical Engineering and Computer Science, University of Ottawa, Ottawa, ON, Canada} 
\IEEEauthorblockA{\IEEEauthorrefmark{2}
    Ericsson Inc., Ottawa, ON, Canada}
Emails: \{mfarz086, hzhan363, melike.erolkantarci\}@uottawa.ca, \{akram.bin.sediq, ali.afana\}@ericsson.com
\vspace{-4ex}
} 

\maketitle

\begin{abstract}
Cell-free Integrated Sensing and Communication (ISAC) aims to revolutionize 6th Generation (6G) networks. By combining distributed access points with ISAC capabilities, it boosts spectral efficiency, situational awareness, and communication reliability.
Channel estimation is a critical step in cell-free ISAC systems to ensure reliable communication, but its performance is usually limited by challenges such as pilot contamination and noisy channel estimates.
This paper presents a novel framework leveraging sensing information as a key input within a Conditional Denoising Diffusion Model (CDDM). In this framework, we integrate CDDM with a Multimodal Transformer (MMT) to enhance channel estimation in ISAC-enabled cell-free systems.
The MMT encoder effectively captures inter-modal relationships between sensing and location data, enabling the CDDM to iteratively denoise and refine channel estimates. 
Simulation results demonstrate that the proposed approach achieves significant performance gains.
As compared with Least Squares (LS) and Minimum Mean Squared Error (MMSE) estimators, the proposed model achieves normalized mean squared error (NMSE) improvements of 8 dB and 9 dB, respectively.
Moreover, we achieve a 27.8\% NMSE improvement compared to the traditional denoising diffusion model (TDDM), which does not incorporate sensing channel information.
Additionally, the model exhibits higher robustness against pilot contamination and maintains high accuracy under challenging conditions, such as low signal-to-noise ratios (SNRs). 
According to the simulation results, the model performs well for users near sensing targets by leveraging the correlation between sensing and communication channels.
\end{abstract}


\vspace{-1ex}
\section{Introduction}

Integrated Sensing and Communication (ISAC) is an innovative approach within the future 6G wireless communication landscape, that aims to unify the traditionally separate functions of communication and radar sensing into a cohesive framework \cite{wei2023integrated}. 
It can enhance the spectral and hardware efficiency in various emerging applications, such as cellular vehicle-to-everything (C-V2X) and extended reality (XR), which simultaneously demand high data rates and precise sensing capabilities.

On the other hand, cell-free multiple-input multiple-output (MIMO) system is a pioneering approach designed to provide seamless and consistent service to all users by eliminating cell boundaries \cite{chen2022survey}. In traditional wireless systems, each user equipment (UE) is connected to a single access point (AP). In contrast, a cell-free MIMO network consists of distributed APs that cooperate to serve users collectively. 
By leveraging this distributed architecture, cell-free MIMO systems ensure consistent signal quality for all users and eliminate inter-cell interference.
The integration of ISAC with cell-free MIMO systems can further enhance both technologies. 
The distributed APs in cell-free networks provide a robust infrastructure for ISAC's sensing and communication functions. In parallel, ISAC's ability to improve situational awareness and spectrum efficiency can optimize the performance of cell-free MIMO systems, particularly in dynamic and dense environments \cite{galappaththige2025cell}.
Additionally, the distributed architecture of cell-free MIMO systems is inherently suited for multi-static sensing, where multiple spatially separated APs work together to transmit and receive sensing signals. 
Multi-static sensing in cell-free MIMO system allows for improved detection, localization, and tracking of objects by leveraging diverse angles.

Channel estimation is crucial for characterizing wireless channel effects, such as fading and interference, to ensure reliable communication and efficient signal processing. However, it faces challenges including noise, time-varying channel, and interference. Additionally, in the cell-free MIMO paradigm, the increasing number of users and distributed AP architecture exacerbates pilot contamination, which arises from the interference of non-orthogonal pilot signals due to the limited availability of pilot sequences \cite{chen2022survey}.
Most existing studies on cell-free ISAC (e.g., \cite{demirhan2024cell,behdad2022power}) assumes perfect channel state information (CSI) for resource allocation and beamforming. However, in practical scenarios, the CSI is often degraded due to noise and pilot contamination. This highlights the need for improved channel estimation techniques to address these impairments in cell-free MIMO systems.

Recently, the application of generative Artificial Intelligence (GAI) algorithms in ISAC-based wireless communication systems has garnered significant interest \cite{wang2024generative, farzanullah2024generative, ghassemi2025generative}.
Generative AI models 
have shown significant potential to enhance the quality of service of 
physical layer of ISAC systems by improving tasks such as channel estimation, signal detection, and beamforming. 
Diffusion model \cite{saharia2022palette} is a GAI algorithm that learns to generate data by iteratively denoising samples from a noise distribution. During training, these models add noise to the data in a controlled manner and then learn to reverse this process to reconstruct the original data.
Diffusion models have demonstrated remarkable success in generating high-quality and diverse data, with notable examples such as Stable Diffusion for image synthesis and DALL-E for text-to-image generation.

Recent studies have explored the application of diffusion models to enhance channel estimation in ISAC wireless systems. One line of research leverages the generative capabilities of diffusion models for synthetic data generation \cite{zhang2024denoising, wang2025generative}. In \cite{zhang2024denoising}, the authors employ a denoising diffusion model (DDM) for synthetic channel generation and demonstrate that the synthetic data can improve sensing channel estimation. Similarly, \cite{wang2025generative} generates synthetic data with reduced noise to aid in the training of Machine Learning (ML) models for the ISAC target detection task. On the other hand, some works have utilized the diffusion model for denoising of sensing target channel estimates \cite{xu2025brownian, yi2024denoising}. The authors in \cite{xu2025brownian} utilize DDM for improving sensing channel estimates, using a virtual channel matrix for prepocessing. 
Likewise, \cite{yi2024denoising} utilized a DDM for improving channel estimation in intelligent reflective surfaces assisted ISAC systems.


Although previous works on ISAC channel estimation primarily focus on enhancing the quality of sensing target channel estimates, they overlook the correlation between sensing target and UE channel estimates, which could enhance UE channel estimation. Moreover, existing works assume the use of orthogonal pilot sequences for channel estimation, and fail to account for pilot contamination, which is a significant challenge in cell-free MIMO systems due to the dense deployment of APs and users.

In this work, we utilize multimodal transformer (MMT)-based Conditional DDM (CDDM) to enhance the quality of user's channel estimation. 
The main contributions of this paper can be summarized as follows:
\begin{itemize}
    \item We propose a novel framework comprising CDDM, designed to enhance the quality of channel estimates at the APs for each user. By leveraging sensing channel information and UE location data, processed through appropriate modality encoders, CDDM effectively learns to denoise the received channel estimates. This approach ensures higher-quality channel estimates, even in the presence of pilot contamination.
    \item To learn the inter-modal relationship between the sensing channel estimates and UE location data, we propose a cross-modal encoder. The cross-modal encoder uses MMT to effectively capture and align the shared information between the two modalities. 
    \item To the best of our knowledge, this is the first work to leverage sensing channel information to enhance uplink UE channel estimation accuracy by exploiting the correlation between the sensing and UE channels.
\end{itemize}
The proposed framework is evaluated on a simulated dataset using the normalized mean squared error (NMSE) metric. The results demonstrate that the proposed model achieves superior performance compared to the conventional Least Square (LS) and Minimum mean-squared error (MMSE) channel estimation techniques, with gains of 8 dB and 9 dB over LS and MMSE, respectively. 
Compared to traditional DDM (TDDM), which does not incorporate sensing estimation information as an input, it achieves an average NMSE improvement of 27.8\%.
Additionally, the model exhibits higher robustness against the pilot contamination issue under challenging conditions. 
It effectively leverages the sensing channel information to denoise the user's channel estimates in close proximity.

The rest of the paper is organized as follows: We first formulate the system model and problem formulation in Section \ref{Sec:SystemModel}. Next, we discuss our proposed ISAC conditioned diffusion model in Section \ref{Sec:Approach}. The simulation settings and results are discussed in Section \ref{Sec:Simulation}. Finally, we conclude the paper in Section \ref{Sec:Conc}.

\section{System Model} \label{Sec:SystemModel}

\begin{figure}
    \centering
    \includegraphics[width=1\linewidth, trim=12pt 10pt 12pt 10pt, clip]{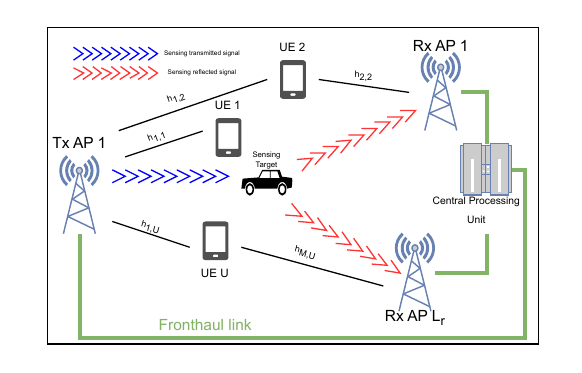}
    \caption{System model}
    \label{fig:systemmodel}
\end{figure}

As shown in Fig. \ref{fig:systemmodel}, we consider a communication system consisting of $L$ APs and $U$ UEs. Each AP is equipped with a millimeter wave (mmWave) uniform linear array (ULA) with $M$ antennas. Meanwhile, each UE is equipped with a single antenna. We consider an ISAC system, where the APs perform both communication and sensing. We also consider a cell-free setting where the APs are connected to a centralized processor for coordination and data processing. This centralized processor enables the APs to collaboratively serve the UEs.
Furthermore, we consider one sensing target, where the UEs are within a maximum distance of $d$ meters from the sensing target. We consider multi-static sensing, where an AP (denoted as $l_t$) is used to transmit the sensing signal, while the remaining $L_r = L - 1$ AP (denoted as $l_r \in \mathcal{L}_r$) act as receivers to collect the reflected signals from the sensing target.
The downlink communication channel from AP $l$ to UE $u$ is indicated by $\mathbf{h}_{lu} \in \mathbb{C}^{M \times 1}$. Furthermore, let the matrix $\mathbf{H^{comm}} \in \mathbb{C}^{L \times U \times M}$ denote the channel from all $L$ APs to all $U$ UEs.

We assume a single-point reflector for the sensing-channel model \cite{demirhan2024cell, behdad2022power}. 
Under this assumption, there exists line of sight (LOS) between the APs and the sensing target. We denote the channel from the transmit AP $l_t$ to the receiver AP $l_r$ through the sensing target as:
\begin{align}
    \mathbf{H}^{sens}_{l_r} = \alpha_{l_r} \mathbf{a}(\theta_{l_r}) \mathbf{a}^H(\theta_{l_t}),
\end{align}
where $\alpha_{l_r} \sim \mathcal{CN}(0, \zeta_{l_t, l_r}^2)$ represents the sensing channel gain, influenced by path loss and radar cross-section (RCS) effects. RCS quantifies an object's ability to reflect radar signals, representing its detectability by radar systems. Meanwhile, $\theta_{l_t} ,\theta_{l_r} \in \mathbb{C}^{M \times 1}$ represents the angle of departure (AoD) to, and angle of arrival (AoA) from the sensing target. For ULA, the steering vector for $\theta_{l_t} ,\theta_{l_r} \sim \mathcal{U}(0, 2\pi)$ is calculated as follows:
\begin{align}
    \mathbf{a}(\theta) = \begin{bmatrix}
1, e^{-j \frac{2\pi d_a}{\lambda} \sin\theta}, \dots, e^{-j \frac{2\pi d_a}{\lambda} (M-1) \sin\theta}
\end{bmatrix}^\mathrm{T},
\end{align}
where $\lambda$ is the carrier's wavelength, and $d_a$ is the distance between adjacent antennas. For simplicity, we consider $d_a = \lambda/2$. 

The communication cycle in a cell-free MIMO architecture comprises of three phases: channel estimation through uplink pilot transmission, uplink data transmission, and downlink data transmission \cite{chen2022survey}. 
We assume that during downlink transmission, some beamforming streams are reserved for sensing capabilities \cite{demirhan2024cell}. Assuming full-duplex capabilites for the APs, the received radar signal $\mathbf{y}_{l_r}$ at the AP $l_r$ can be expressed as \cite{demirhan2024cell}:
\begin{align}
    \mathbf{y}_{l_r} = \alpha_{l_r} \mathbf{a}(\theta_{l_r}) \mathbf{a}^H(\theta_{l_t}) \mathbf{x}_{l_t} + \mathbf{n}_{l_r},
\end{align}
where $\mathbf{n}_{l_r}$ is the noise at the receiver of AP $l_r$. Assuming knowledge of $\mathbf{x}_{l_t}$ at the APs, the receiver AP $l_r$ can estimate the sensing channel $\mathbf{H}^{\text{sens}}_{l_r}$. Furthermore, we adopt the Swerling-I model \cite{richards2010principles}, which assumes a static sensing channel from downlink transmission to subsequent uplink channel estimation. 
This work focuses on improving the channel estimation for $\mathbf{H}^{\text{comm}}$ during the uplink pilot transmission phase using sensing channel estimates $\mathbf{H}^{\text{sens}}_{l_r}, \{l_r \in \mathcal{L}_r\}$, and the UE location data. Let $\mathbf{k}_u\in\mathbb{R}^2$ denote the position of user $u$. We assume that the UE transmits their location data during uplink transmission.

We assume that during the uplink channel estimation stage, the UEs are assigned $\tau_p$ orthogonal pilots for transmission, each of length $\tau_p$. In most systems, $\tau_p < U$, and hence pilot symbols must be reused among different UEs. Let $s_u \in \{1, \dots, \tau_p\}$ denote the pilot assigned to user $u$, and let $\mathcal{S}_u$ represent the set of UEs that share the pilot $s_u$. When the UEs transmit their pilots simultaneously, the signal received at AP $l$ after despreading is given by \cite{chen2022survey}:
\begin{align}
    \mathbf{y}^{\text{pilot}}_{s_u,l} = \sum_{i \in \mathcal{S}_u} \sqrt{\tau_p p_i} \mathbf{h}_{il} + \mathbf{n}_l,
\end{align}
where $p_i$ denotes the transmit power of UE $i$, $\mathbf{h}_{il}$ is the channel vector between UE $i$ and AP $l$, and $\mathbf{n}_l$ represents the noise at the receiver of AP $l$. Further, $\mathbf{h}_{il}$ is the uplink channel, that is the reciprocal of the downlink channel $\mathbf{h}_{li}$ in the time division duplexing system. 
If $\tau_p \ge U$, there would be no interference between UEs as no two UEs would share the same pilot. However, in most cell-free paradigms $\tau_p < U$, which results in pilot contamination. This effects the quality of the channel estimates received, degrading the performance of wireless tasks such as beamforming. This degradation occurs because accurate channel estimates are critical for optimizing beamforming, as they directly influence the alignment and efficiency of signal transmission and reception.
The LS estimate is commonly used to estimate the channel by minimizing the squared error between the received pilot signal and the expected signal. Let $\mathbf{h}^{LS}_{lu}\in\mathbb{C}^{M\times1}$ denote the estimated LS downlink channel by the AP $l$ for UE $u$, and $\mathbf{H}^{LS}\in\mathbb{C}^{L\times U \times M}$ denote the matrix for LS channel estimated for all AP-UE pairs.

\subsection{Problem Formulation}

The objective of the work is to denoise the estimated LS channel $\mathbf{H}^{LS}$ conditioned on the sensing channel estimate $\mathbf{H}^{\text{sens}}_{l_r}, \{l_r \in \mathcal{L}_r\}$ and the UE location data $\mathbf{k}_u$. Let $\mathbf{H}^{den}$ denote the denoised output of our model. 
So we can formulate the problem as:
\begin{align}
\min_{\mathbf{H}^{\text{den}}} & \quad \|\mathbf{H}^{\text{den}} - \mathbf{H}\|_F^2 \\
\text{s.t.} & \quad 
\begin{cases}
\mathbf{H}^{\text{den}} = f\left(\mathbf{H}^{\text{LS}}, \{\mathbf{H}^{\text{sens}}_{l_r}\}_{l_r \in \mathcal{L}_r}, \mathbf{k}_u\right), \\
\mathbb{E}[|\mathbf{x}_{l_t}|^2=1], \\
p_i \leq P_{max},
\end{cases}
\end{align}
where $\mathbf{H}$ is the true channel information, $F$ represents the Frobenius norm, and $P_{max}$ is the maximum transmit power for the UEs.

According to the problem formulation, the goal of this work is to refine the estimated noisy channel $\mathbf{H}^{LS}$ by leveraging sensing-based channel estimation and UE location data. The objective is to produce a denoised channel estimate, $\mathbf{H}^{den}$, that closely approximates the true channel $\mathbf{H}$.
To solve this problem, we utilize a conditional denoising diffusion model to achieve the objective. The proposed MMT-based CDDM effectively captures the complex relationships between the noisy channel estimates, sensing data, and UE location information, enabling accurate reconstruction of the true channel. The details of MMT-based CDDM is introduced in the next section.

\section{Multimodal Transformer-based ISAC conditioned Denoising Diffusion Model
} \label{Sec:Approach}

\begin{figure*}
    \centering
    \includegraphics[width=1\linewidth, trim=20pt 6pt 20pt 6pt, clip]{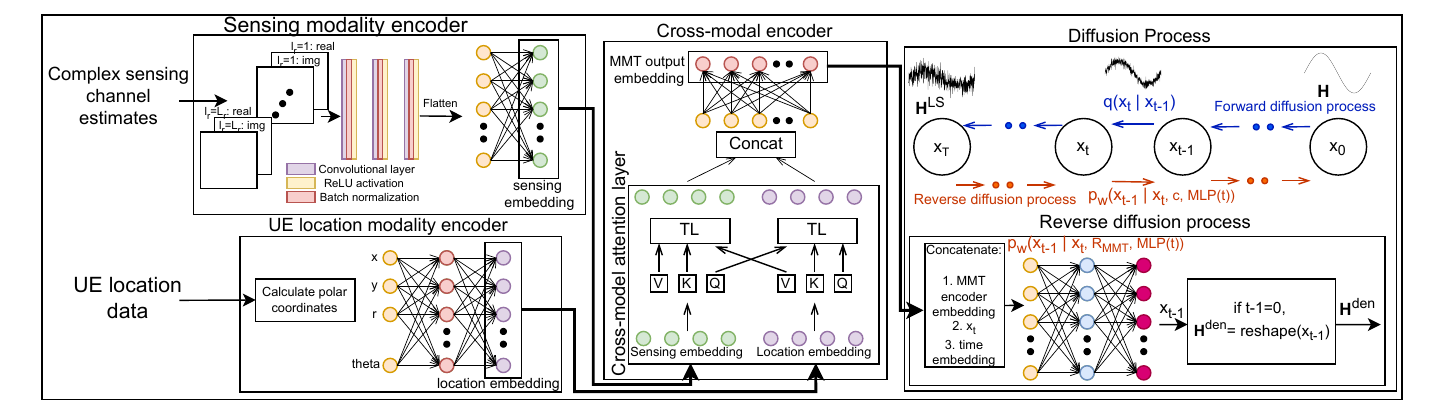}
    \caption{The architecture of the proposed approach is as follows: the input modalities are first processed through modality-specific encoders, producing embeddings that are subsequently passed through a cross-modal encoder to capture inter-modal relationships. These learned relationships serve as a conditioning mechanism for the diffusion process, which is designed to denoise the LS estimated channel and generate refined channel estimates.
    }
    \label{fig:proposedapproch}
\end{figure*}

The proposed CDDM framework is shown in Fig. \ref{fig:proposedapproch}. The framework reconstructs the original channel $\mathbf{h}_{u}$ for each UE $u$ for all APs $L$ from the LS estimate $\mathbf{h}^{LS}_{u}$.
The framework includes three major components: modality encoders, cross-modal encoder, and diffusion process.
We will explain these steps below. Note that, the denoising diffusion model is integrated with cross-modal encoder, based on multimodal transformer, to enhance the understanding of complex data across diverse input modalities, such as sensing channel information and UE location data. This integration exploits the strength of diffusion models in progressively refining noisy inputs through iterative denoising steps, enabling the reconstruction of highly accurate channel estimates even under challenging noise conditions.

\subsection{Modality encoders}
We first use specially designed modality encoders to process sensing channel estimates, $\{\mathbf{H}^{\text{sens}}_{l_r}\}_{l_r \in \mathcal{L}_r}$, and the UE location data $\mathbf{k}_u$, for generating embeddings to train our model.
The encoders play a critical role in extracting meaningful features from the input data, ensuring that the embeddings capture the essential characteristics required for effective model training and accurate predictions. The modality encoders are introduced as follows:

\subsubsection{Complex sensing estimate encoder}
The sensing channel estimate is represented as a complex matrix. To process this matrix, the real and imaginary components for each $L_r$ APs were separated into distinct channels. Consequently, the input consisted of $2 \times L_r$ channels, each with a shape of $M \times M$. These inputs were then processed through $C$ convolutional neural network (CNN) layers. Each layer $i$ comprised a convolutional layer with $c_i$ filters, followed by batch normalization and a rectified linear unit (ReLU) activation function. The resulting output was flattened and passed through a fully connected linear neural network (NN) layer to produce an embedding vector $\mathbf{R}_s$. This approach allows the model to effectively capture both spatial and spectral features of the sensing channel, improving its ability to learn meaningful representations.

\subsubsection{UE location encoder}
The $(x, y)$ coordinates of the UE were used to calculate the polar coordinates $(r, \theta_{lu})$, where $r$ represents the distance to the AP, and $\theta_{lu}$ denotes the angle. These four inputs were concatenated and passed through a series of linear NN layers to produce an embedding $\mathbf{R}_{loc}$. This transformation allows the model to utilize both spatial and angular information, improving its capacity to learn informative representations.

\subsection{Cross-modal Encoder}
The cross-modal encoder utilizes MMT, and enables our model to learn the inter-modal relationships between diverse input modalities. 
Self-attention helps the model capture long-range dependencies within a single modality, enhancing its ability to extract meaningful features. The cross-modal attention mechanism allows tokens from one modality to attend to tokens from another modality, facilitating the integration of different input data. In this work, we first pass the sensing estimate embeddings through a self-attention layer. Subsequently, a cross-modal attention layer is applied between the two embeddings. The resulting output is then processed through a feedforward network to produce an embedding vector $\mathbf{R}_{MMT}$.

\subsection{Conditional Denoising Diffusion Model}

Denoising diffusion 
models are a class of GAI models that iteratively remove noise from corrupted data through a Markovian diffusion process. In this work, we employ CDDM \cite{ho2022classifier} that leverages sensing and location embeddings as conditioning inputs, enabling the model to reconstruct channel estimates with higher accuracy.

\subsubsection{Forward Diffusion Process}
The forward diffusion process gradually corrupts the clean channel estimate $\mathbf{x}_0$ by adding Gaussian noise over $T$ time steps according to a predefined noise schedule. At each step, the noisy sample $\mathbf{x}_t$ is generated from $\mathbf{x}_{t-1}$ as:
\begin{equation} \label{Eq:forwardDiff}
q(\mathbf{x}_t | \mathbf{x}_{t-1}) = \mathcal{N}(\mathbf{x}_t; \sqrt{\alpha_t} \mathbf{x}_{t-1}, (1 - \alpha_t) \mathbf{I}),
\end{equation} 
where $\alpha_t$ is a time-dependent scaling factor that controls the amount of noise added at each step. The noise schedule is designed such that the noise addition is small in the initial steps and increases progressively in later steps.

Using the Markov property of the forward process, the cumulative noise added to a clean sample $\mathbf{x}_0$ at time step $t$ can be expressed as:
\begin{equation}
q(\mathbf{x}_t | \mathbf{x}_0) = \mathcal{N}(\mathbf{x}_t; \sqrt{\bar{\alpha}_t} \mathbf{x}_0, (1 - \bar{\alpha}_t) \mathbf{I}),
\end{equation}
where $\bar{\alpha}_t = \prod_{s=1}^{t} \alpha_s$ is the cumulative product of the noise scaling factors up to time step $t$. This formulation allows direct sampling of $\mathbf{x}_t$ from $\mathbf{x}_0$ without iterating through all intermediate steps.

\subsubsection{Reverse Process and Model Architecture}
The reverse diffusion process aims to iteratively reconstruct the clean sample $\mathbf{x}_0$ from a noisy sample $\mathbf{x}_T$ by predicting the intermediate states \(\mathbf{x}_{t-1}\) from \(\mathbf{x}_t\). The reverse process is modeled as a parameterized Gaussian distribution:
\begin{equation}
p_w(\mathbf{x}_{t-1} | \mathbf{x}_t) = \mathcal{N}(\mathbf{x}_{t-1}; \mu_w(\mathbf{x}_t, \tau_t, R_{\text{MMT}}), \sigma_t^2 \mathbf{I}),
\end{equation}
where $\mu_w$ is the predicted mean of the reverse process, $\sigma_t^2$ is the variance, and $\mathbf{R}_{\text{MMT}}$ is the conditioning input. The conditioning input $\mathbf{R}_{\text{MMT}}$ is the output of the cross-modal encoder, which fuses the sensing and location embeddings to provide additional context for the denoising process. The time step $t$ is encoded as an embedding $\tau_t$ to inform the model of the current diffusion step.

The neural network used for denoising is a fully connected multi-layer perceptron (MLP) that predicts the next state $\mathbf{x}_{t-1}$ directly from the current noisy input $\mathbf{x}_t$:
\begin{equation}
\mathbf{x}_{t-1} = f_w(\mathbf{x}_t, \tau_t, R_{\text{MMT}}) = \text{MLP}(\mathbf{x}_t, \tau_t, \mathbf{R}_{\text{MMT}}),
\end{equation}
where $w$ represents the model parameters, and the MLP consists of multiple linear layers. The inputs to the MLP include the flattened real and imaginary parts of the noisy sample $\mathbf{x}_t$, the time embedding $\tau_t$, and the conditional embedding $\mathbf{R}_{\text{MMT}}$.

During inference, the reverse process is run iteratively, starting from a noisy sample $\mathbf{x}_T$. At each step, the model predicts $\mathbf{x}_{t-1}$ from $\mathbf{x}_t$, and this process is repeated until $t = 0$, at which point the final reconstructed clean sample $\mathbf{x}_0$ is obtained.

\subsubsection{Training Objective}
The model is trained to predict the next state $\mathbf{x}_{t-1}$ from the noisy input $\mathbf{x}_t$ by minimizing the following NMSE loss:
\begin{equation}
\mathcal{L}_{\text{NMSE}} = \mathbb{E}_{\mathbf{x}_0, \epsilon, t} \left[ \frac{\| \mathbf{x}_{t-1} - \hat{\mathbf{x}}_{t-1} \|^2}{\| \mathbf{x}_{t-1} \|^2} \right],
\end{equation}
where $\epsilon$ represents the noise added during the forward process, and $\hat{\mathbf{x}}_{t-1}$ is the model's prediction of the next state. The NMSE loss ensures that the model learns to accurately predict the intermediate states while being robust to variations in the scale of the data.


\section{Simulation Setting and Results} \label{Sec:Simulation}

We consider an outdoor urban environment with dimensions of $100 \times 100 \, \text{m}^2$. The scenario includes one ISAC multi-static transmitting AP, located at the coordinate $(0, 50)$. Additionally, the number of receiving multi-static APs, denoted as $L_r$, is set to 2. The receiving APs are randomly positioned along the y-axis with a fixed x-coordinate of $x = 100$. All three APs participate in communication with the UEs. 
Each AP is equipped with a ULA 
consisting of $M = 8$ antennas. The position of the sensing target is randomly generated. The number of UEs, denoted as $U \in [3,9]$, and the number of pilot symbols, $\tau_p \in \{4,8\}$, are varied during the simulations.
The UEs are allowed to be at a maximum distance $d \in [2.5, 20]$ from sensing target. Furthermore, we vary the UE pilot transmit power $p_u$, such that the SNR at the AP is in the range $0-10 \text{ dB}$. 
Moreover, we adopt the 3GPP Urban Mobility (UMi) model for path loss and shadowing \cite{3gpp_tr_38_901}, given by:
\begin{align}
    \text{PL (dB)} = 22.4 + 35.3 \log_{10} (\text{dist}) + 21.3 \log_{10} (f_c) + \mathcal{X},
\end{align}
where $f_c$ is the career frequency that is set as 28 GHz, and $\mathcal{X} \sim \mathcal{N}(0, 7.82)$ is the shadowing effect. Moreover, we consider Rician fading with a K-factor of 10.

We generated a dataset consisting of 10,000 samples. 
The dataset was divided into 60\%, 20\%, and 20\% for the training, validation, and test sets, respectively. The training set was used to train our proposed model, referred to as ISAC-CDDM. 
The trained model was subsequently evaluated on various diverse scenarios to quantify its effectiveness across different conditions.

In the proposed model, as depicted in Fig. \ref{fig:proposedapproch}, we adopt 3 convolutional layers in the sensing modality encoder with $\{16, 32, 64\}$ filters. The output from the third convolutional block is flattened, and passed through a linear NN layer to generate embedding of size $|\mathbf{R}_s| = 16$.
Meanwhile, the NN of UE location encoder consists of two layers in total: one hidden layer with 64 neurons and one output layer with size $|\mathbf{R}_{loc}|=16$. The sensing modality embedding $\mathbf{R}_s$, and the UE location embedding $\mathbf{R}_{loc}$ are passed through a MMT encoder consisting of 1 layer with 8 multihead attention mechanisms. The MMT encoder layer consists of self-attention layer, cross-attention layer, and the NN that transforms the output to an embedding of size $|\mathbf{R}_{MMT}| = 128$. 
For the diffusion process, the $\alpha_t$ in Eq. (\ref{Eq:forwardDiff}) is reduced from 0.9999 to 0.98 across the diffusion steps. The MLP for reverse diffusion process consists of two hidden layers, with 512 neurons in each.
The model is trained using the RMSprop optimizer with a batch size of 32. The initial learning rate is set to 0.001 and is reduced by a factor of 0.5 if the validation loss does not improve over a span of five consecutive epochs. Training is conducted for a maximum of 200 epochs. The model was developed using PyTorch 2.2.0 and trained on a GPU. The training process required approximately one hour to complete.

We denote our proposed algorithm as ISAC-CDDM, and use the following three benchmarks for comparison:
\begin{enumerate}
    \item LS estimator: The LS estimator minimizes the error between the observed and estimated signals without requiring prior knowledge of the channel or noise. It is simple and efficient but less accurate in noisy conditions.
    \item MMSE estimator: The MMSE estimator minimizes the mean square error using prior knowledge of channel and noise statistics. It is more accurate than LS estimator in noisy environments but computationally complex due to matrix inversions and the need for statistical information.
    \item Traditional Denoising Diffusion Model (TDDM): The TDDM model is trained without UE location data or sensing information. This serves as a benchmark to assess performance without sensing channel information as an input.
\end{enumerate}

To evaluate the effectiveness of our model, we use the normalized mean squared error (NMSE), that is given by:
\begin{align}
    NMSE (\mathbf{H}^{den},\mathbf{H}) = \mathbb{E}\left[\frac{||\mathbf{h}_{lu}^{den}-\mathbf{h}_{lu}||^2}{||\mathbf{h}_{lu}||^2}\right]
\end{align}
NMSE accounts for the scale of the true channel, providing a fair comparison by normalizing the error relative to the channel's power.


\begin{figure}
    \centering
    \includegraphics[width=0.85\linewidth, trim=20pt 13pt 50pt 30pt, clip]{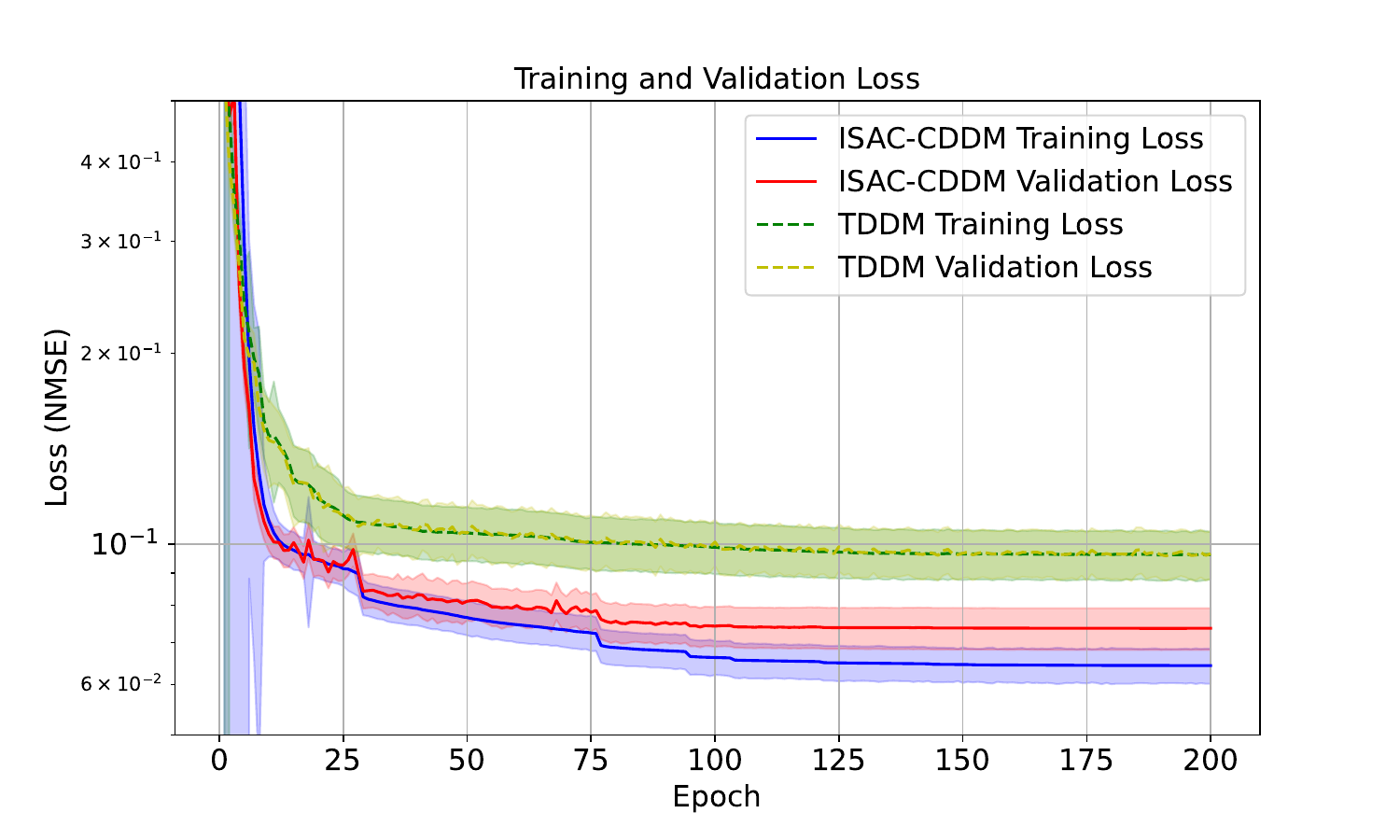}
    \caption{Training and validation loss during training process.}
    \label{fig:trainperf}
\end{figure}
Fig. \ref{fig:trainperf} shows the training and validation loss of our model and TDDM during the training phase. Our model achieves lower training loss than TDDM, showing improved fit to training data. The sharp initial loss decrease shows rapid learning of data structure and noise, while the plateau indicates convergence and steady optimization.

\begin{figure}
    \centering
    \includegraphics[width=1\linewidth, trim=8pt 6pt 30pt 33pt, clip]{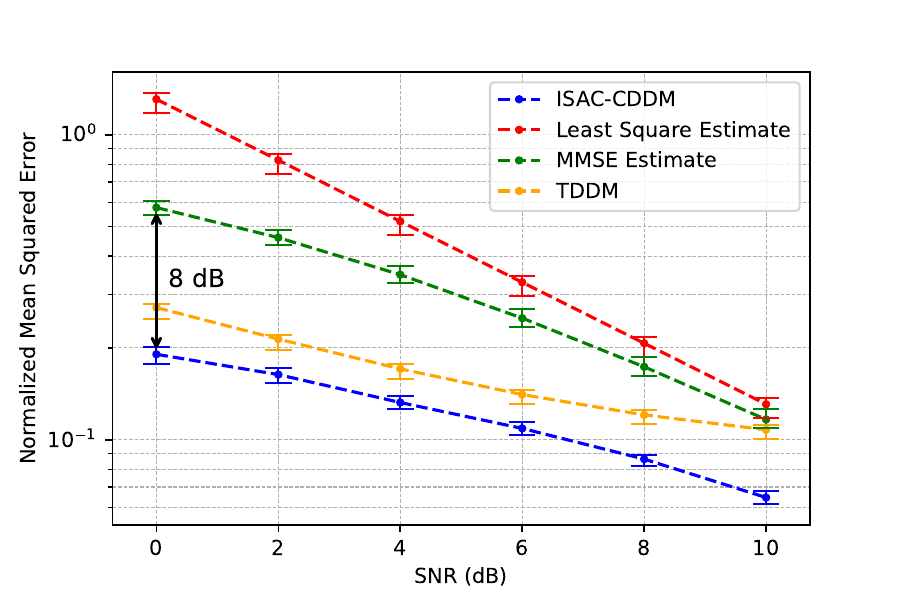}
    \caption{NMSE against received SNR (dB), with $U=8, \tau_p=8, d=10$ m.}
    \label{fig:nmse_snr}
\end{figure}

For Fig. \ref{fig:nmse_snr}, we change the SNR received at AP by adjusting the UE transmit power. We set $U=8$, $\tau_p = 8$, and $d = 10 \, \text{m}$. This represents an idealized scenario where all UEs have orthogonal pilots, resulting in no pilot contamination. Our ISAC-CDDM outperforms other benchmarks. Specifically, our model provides a gain of 8 dB and 9 dB compared to the LS and MMSE estimators, respectively, in the low SNR region. The ISAC-CDDM also outperforms other benchmarks in the high SNR region.
Furthermore, there is an average 27.8\% NMSE improvement as compared to TDDM.

\begin{figure}
    \centering
    \vspace{-1ex}
    \includegraphics[width=1\linewidth, trim=8pt 6pt 30pt 33pt, clip]{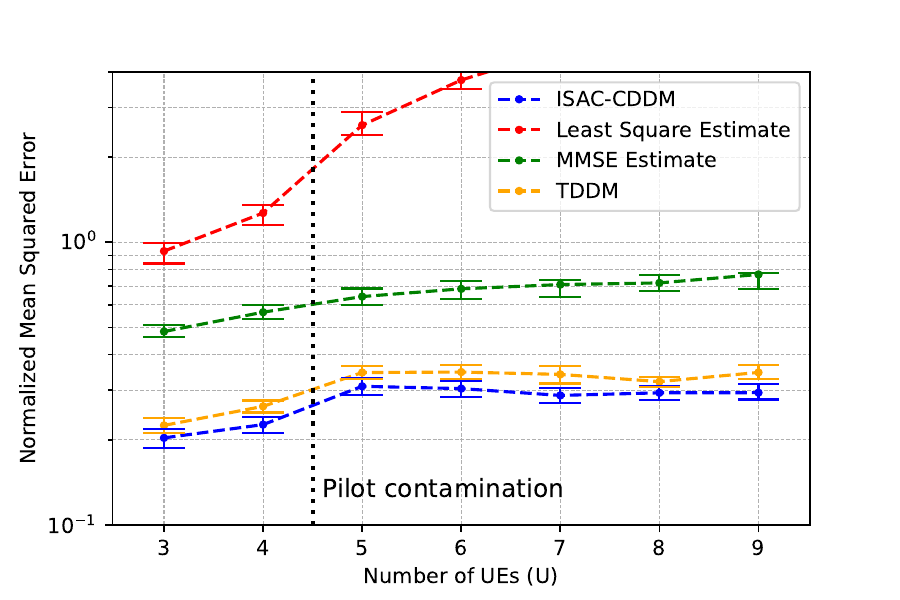}
    \caption{NMSE on increasing number of UEs $U$ with $\tau_p = 4, d =10$ m, SNR = $0$ dB.
    The NMSE for the LS method for $U>6$, increases significantly. For the sake of figure clarity, we do not include these results in the plot.
    }
    \label{fig:nmse_ue}
\end{figure}

Fig. \ref{fig:nmse_ue} illustrates the effect of pilot contamination by increasing the number of UEs, $U$, while keeping the number of pilots constant at $\tau_p = 4$. The model is further evaluated under challenging conditions with an SNR of $0$ dB, where the signal and noise power are equal. As expected, the performance of the LS estimator declines drastically for $U > 4$, due to pilot contamination. Meanwhile, the MMSE estimator is less affected, as it is more resistant to pilot contamination. It leverages statistical knowledge of the channel and noise, enabling it to distinguish between desired and interfering signals. Our ISAC MT-DDM outperforms all benchmarks, demonstrating that it is least affected by the pilot contamination problem.

\begin{figure}
    \centering
    \includegraphics[width=1\linewidth, trim=0pt 6pt 2pt 5pt, clip]{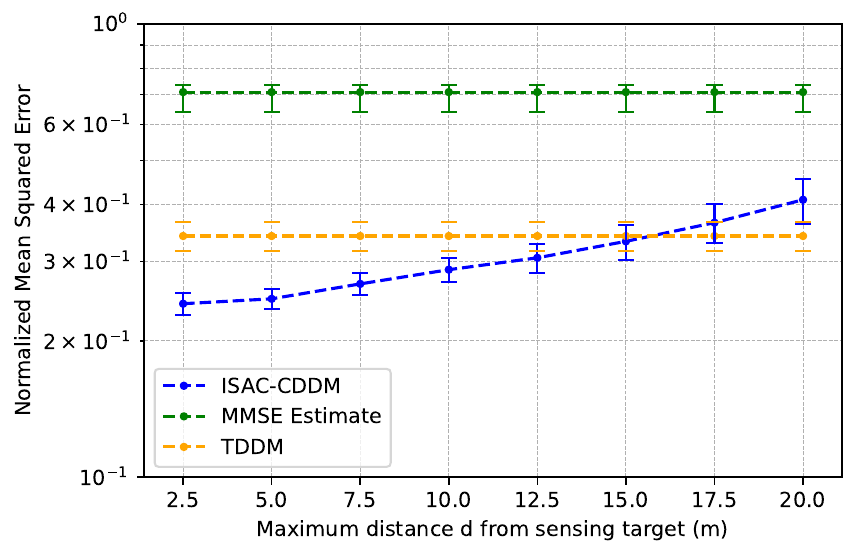}
    \caption{NMSE on increasing minimum distance $d$ with $\tau_p = 4, U =8$, SNR = $0$ dB. 
    The LS method is omitted for the sake of clarity.
    }
    \label{fig:nmse_dist}
\end{figure}

Fig. \ref{fig:nmse_dist} illustrates the impact of increasing the minimum distance, $d$, of the UEs from the sensing target, under pilot contamination and SNR $= 0$ dB. The NMSE of the benchmarks remain constant as they are independent of the sensing target. Meanwhile, the NMSE of our ISAC-CDDM model increases with $d$, and is lower than TDDM for $d \leq10$ m.  This indicates that the correlation between sensing channel information and UE channel estimates decreases as UEs move farther from the sensing target. 

\section{Conclusion} \label{Sec:Conc}

This work proposes a novel approach that incorporates a multimodal transformer-based conditional denoising diffusion model to reconstruct true channels from noisy channel estimates at the APs in a cell-free ISAC-based 6G network. Specifically, the approach uses ISAC sensing channel estimates and UE location data as conditions for the diffusion model to denoise channel estimates. Our experimental results show that, compared to MMSE and LS channel estimation techniques, we achieve gains of 8 dB and 9 dB, respectively. Furthermore, our model effectively counters the pilot contamination issue, even under challenging conditions with low SNR. The sensing target channel estimates serve as a valuable input to improve the channel estimates of users in close proximity.

\section*{Acknowledgement}
This work has been supported by MITACS and Ericsson Canada.
\vspace{-1ex}
\balance
\bibliographystyle{IEEEtran} 
\bibliography{pimrc_paper}
\balance \balance


\end{document}